\documentclass{template/egpubl}
\usepackage{template/sca2022}
\SpecialIssuePaper
\usepackage[T1]{fontenc}
\usepackage{template/dfadobe}  
\biberVersion
\BibtexOrBiblatex
\usepackage[backend=biber,bibstyle=template/EG,citestyle=alphabetic,backref=true]{biblatex}
\electronicVersion
\PrintedOrElectronic
\usepackage[pdftex]{graphicx}
\usepackage{template/egweblnk}

\usepackage{amsmath}
\usepackage{amsfonts}
\usepackage{multirow}
\usepackage{subcaption}
\usepackage{xspace}
\usepackage{relsize}

\newcommand{\UnderPressureBox}[3]{%
    \raisebox{#1}{#3}\kern #2%
}
\newcommand{\UnderPressure}{%
    \textsc{%
        \UnderPressureBox{-0.42em}{-0.10em}{\larger[2]U\smaller[2]}%
        \UnderPressureBox{0.0em}{-0.10em}{n}%
        \UnderPressureBox{0.0em}{-0.08em}{d}%
        \UnderPressureBox{0.0em}{-0.08em}{e}%
        \UnderPressureBox{0.0em}{0.0em}{r}%
        \UnderPressureBox{-0.42em}{-0.08em}{\larger[2]P\smaller[2]}%
        \UnderPressureBox{0.0em}{-0.11em}{r}%
        \UnderPressureBox{0.0em}{-0.05em}{e}%
        \UnderPressureBox{0.0em}{-0.05em}{s}%
        \UnderPressureBox{0.0em}{-0.08em}{s}%
        \UnderPressureBox{0.0em}{-0.12em}{u}%
        \UnderPressureBox{0.0em}{-0.11em}{r}%
        \UnderPressureBox{0.0em}{0.0em}{e}%
    }\xspace%
}

\newcommand{\UnderPressureWithUrl}{%
   \href{%
        https://github.com/InterDigitalInc/UnderPressure%
    }{\color{black}{%
        \UnderPressure%
    }}%
}

\newcommand{\eg}{e.g.\xspace}
\newcommand{\Eg}{E.g.\xspace}
\newcommand{\ie}{i.e.\xspace}
\newcommand{\etal}{et~al.\xspace}
\newcommand{\dd}{$2D$\xspace}
\newcommand{\ddd}{$3D$\xspace}

\usepackage{tikz}
\usetikzlibrary{positioning}

\title[%
    Deep Learning for Foot Contact Detection, Ground Reaction Force Estimation and Footskate Cleanup%
]{%
    UnderPressure: Deep Learning for Foot Contact Detection, Ground Reaction Force Estimation and Footskate Cleanup%
}

\author[%
    L. Mourot, L. Hoyet, F. Le Clerc~\& P. Hellier%
]{\parbox{\textwidth}{\centering%
    Lucas Mourot$^{1,2}$\orcid{0000-0001-8441-892X}, %
    Ludovic Hoyet$^{1}$\orcid{0000-0002-7373-6049}, %
    Fran{\c{c}}ois Le Clerc$^{2}$\orcid{0000-0003-0519-8581} and %
    Pierre Hellier$^{2}$ \orcid{0000-0003-3603-2381}%
}\\{\parbox{\textwidth}{\centering%
    $^1$Inria, Univ Rennes, CNRS, IRISA\\
    $^2$InterDigital, Inc%
}}}

\volume{41}       
\issue{8}         
\pStartPage{1}    

\begin{document}
    
    \teaser{
    \centering
    \includegraphics[width=0.91\linewidth]{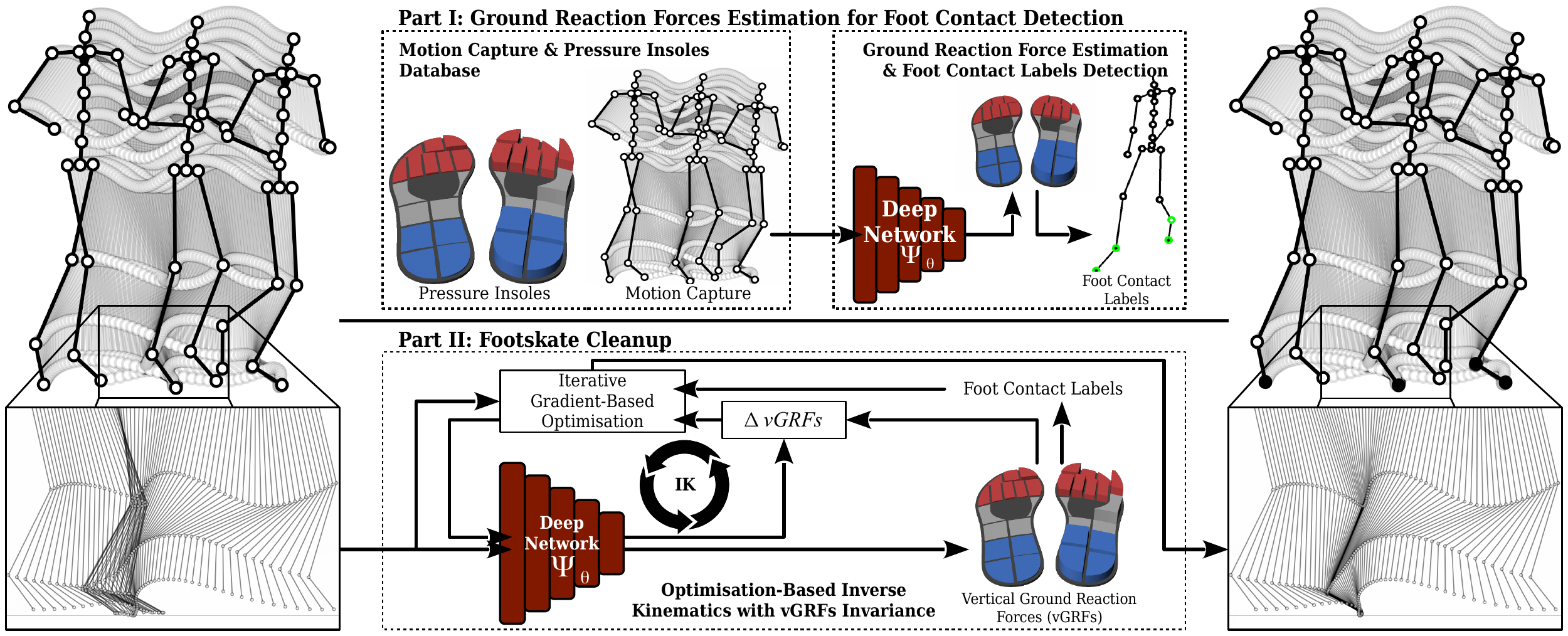}
    \caption{
        We leverage \UnderPressure, a novel publicly available dataset of motion capture synchronized with pressure insoles data, to estimate vertical Ground Reaction Forces (vGRFs) from motion data with a deep neural network and derive foot contact labels. We further clean up footskate artefacts through an optimisation-based inverse kinematics algorithm while enforcing vGRFs invariance.
    }
  \label{fig:teaser}
}
    \maketitle
    \begin{abstract}
    Human motion synthesis and editing are essential to many applications like video games, virtual reality, and film post-production. However, they often introduce artefacts in motion capture data, which can be detrimental to the perceived realism. In particular, footskating is a frequent and disturbing artefact, which requires knowledge of foot contacts to be cleaned up. Current approaches to obtain foot contact labels rely either on unreliable threshold-based heuristics or on tedious manual annotation.
    In this article, we address automatic foot contact label detection from motion capture data with a deep learning based method. To this end, we first publicly release \UnderPressure, a novel motion capture database labelled with pressure insoles data serving as reliable knowledge of foot contact with the ground. Then, we design and train a deep neural network to estimate ground reaction forces exerted on the feet from motion data and then derive accurate foot contact labels. The evaluation of our model shows that we significantly outperform heuristic approaches based on height and velocity thresholds and that our approach is much more robust when applied on motion sequences suffering from perturbations like noise or footskate.
    We further propose a fully automatic workflow for footskate cleanup: foot contact labels are first derived from estimated ground reaction forces. Then, footskate is removed by solving foot constraints through an optimisation-based inverse kinematics (IK) approach that ensures consistency with the estimated ground reaction forces. Beyond footskate cleanup, both the database and the method we propose could help to improve many approaches based on foot contact labels or ground reaction forces, including inverse dynamics problems like motion reconstruction and learning of deep motion models in motion synthesis or character animation. Our implementation, pre-trained model as well as links to database can be found at \href{https://github.com/InterDigitalInc/UnderPressure}{github.com/InterDigitalInc/UnderPressure}.

\end{abstract}
        
    \section{Introduction}
        \label{section:Introduction}
Perceived realism is central to human animation; however, artefacts are often introduced whenever editing or synthesising motion data. These include feet sliding on, passing through or floating above the ground, which are known to disturb human perception even at very low intensities \cite{ref:PrHO11}. In this paper we propose a fully automatic approach to foot contact detection and footskate artefacts removal.

As of today, foot contacts are derived from motion sequences using simple heuristics, most commonly hand-crafted thresholds over velocity and proximity relative to the ground. These approaches suffer from three main limitations. First, the terrain height map must be known, meaning that these approaches are most of the time not applicable to uneven or inclined ground. Second, optimal thresholds are not universal and must be manually tuned, ideally for every type of motion, morphology and contact location (\eg heel or toe). Finally, even optimal thresholds fail at accurately detecting every foot contact, which implies that tedious manual checking is necessary when using these approaches. Finally, even optimal thresholds are far from being $100\%$ accurate and lack of robustness, which implies that tedious manual checking is necessary when using these approaches.

In this paper, we propose a data-driven foot contact detection method from motion that outperforms traditional heuristic approaches. First, we introduce \UnderPressure, a novel publicly available dataset composed of diverse motion capture data from 10 individuals, synchronised with pressure insoles data, from which corresponding vertical Ground Reaction Forces (vGRFs) and foot contact labels can be derived. Then, our key idea is to model vGRFs with a deep neural network, providing a finer representation of interactions of the feet with the ground than binary contact labels. Furthermore, vGRFs are related to the dynamics of motion unlike binary labels. To this end, we train a deep neural network to estimate the distribution of vGRFs over the feet, from which foot contact labels can be calculated, encouraging robustness and generalisability since the network needs a relatively high-level understanding of the motion dynamics to estimate accurate vGRFs.

We then provide a quantitative evaluation of our model against an \emph{optimal thresholds (OT)} baseline that relies on thresholding the velocity and height above ground of foot joints, linear and non-linear learned generalisations of the OT baseline, as well as an ablative study of the proposed architecture. We further experiment how these models behave in different challenging conditions representative of real-world applications.

Finally, we demonstrate the generalisability of our approach on motion sequences from other databases, as well as its integration in a fully automatic footskate cleanup workflow. The novelty of this workflow is to leverage our deep neural network by enforcing invariance between vGRFs estimated from input and optimised motion sequences to better guide the IK optimisation and maintain the consistency of interactions between feet and ground.

To the best of our knowledge, we propose the first method of foot contact detection for animation learned on a significant amount of diverse motion data labelled with foot pressures. Here is a summary of our key contributions:
\begin{itemize}
    \item a unique and publicly available motion capture dataset, called \UnderPressure, composed of 5.6 hours of diverse motions from 10 individuals and synchronized with pressure insoles data,
  
    \item a deep neural network modelling part of the motion dynamics from joint positions, \ie vGRF distribution over the feet,
    
    \item a robust state-of-the-art method for foot contact detection, providing about $95\%$ correct foot contact labels at $100~Hz$ on new subjects,
    
    \item a fully automatic workflow for footskate cleanup built on top of our foot contact detection method, preserving vGRFs estimated by our deep neural network.
    
\end{itemize}

    \section{Related Work}
        \label{section:RelatedWork}
Early works in human animation already considered kinematic constraints such as foot contacts \cite{ref:BiBa98, ref:LCRH02, ref:KoSG02, ref:IkAF06, ref:LeBo06}. Foot contact labels are helpful in numerous applications: they are often necessary to clean up foot artefacts such as foot sliding \cite{ref:KoSG02}, \eg by enforcing foot constraints via Inverse Kinematics (IK) \cite{ref:MiCh12, ref:HoKS17, ref:SZKS19, ref:ALLS20, ref:SZKZ20, ref:HYNP20}; likewise, they are required to quantify such foot artefacts, \eg for evaluation purposes \cite{ref:SZKS19, ref:SZKZ20, ref:ZSKS18, ref:LZCv20}; they are also helpful to mitigate the well-known problem of mean collapse / drift away in human motion prediction, particularly with deterministic and recurrent models \cite{ref:HeAB20, ref:MaBR17, ref:WaCX19, ref:HYNP20}, to disambiguate human motion modelling \cite{ref:XWCC15, ref:HoSK16, ref:PaGA18, ref:PFAG20, ref:HoKS17}, and more generally to leverage contact-based loss functions for increased quality and robustness \cite{ref:HoKS17, ref:LeLL18, ref:WHSZ19, ref:PRLL19, ref:SAAK20}; foot contact information is also deeply involved in character control based on physical simulation \cite{ref:ALXP19, ref:KwLv20, ref:WoLe19, ref:XLKv20, ref:YuTL18} where ground reaction forces are explicitly modelled in the physics engine, and also relates to foot-placement strategies that are a real challenge for locomotion policies \cite{ref:Peva17}. 
In the following, we overview existing approaches for foot contacts labels detection (\ref{section:RelatedWork:Contacts}) and ground reaction forces estimation (\ref{section:RelatedWork:GRFs}), as well as existing databases of motion data labelled with information on foot contacts (\ref{section:RelatedWork:Databases}).

\subsection{Foot Contact Labels Detection} 
\label{section:RelatedWork:Contacts}
    The most widespread family of approaches in both animation research and industry to extract foot contact labels from motion data relies on simple heuristics with hand-crafted thresholds, applied to velocity and proximity relative to the ground as proposed by Bindiganavale and Badler~\cite{ref:BiBa98} or Lee~\etal~\cite{ref:LCRH02}. However, foot contact identification methods relying on heuristics based on joint position and velocity lack temporal precision and are not reliable~\cite{ref:LeBo06}. Therefore, they are either limited to algorithms that are insensitive to the accuracy of contact labels \cite{ref:PRLL19}, or require tedious manual checking and corrections \cite{ref:WaCX19, ref:MSZB18, ref:HoKS17} to produce faithful contact detection results. Moreover, the accuracy of these approaches often dramatically drops when decreasing motion sequences quality, e.g. increasing noise or artefacts.
    
    Other heuristic approaches have been sparsely investigated, including the work of Le Callennec and Boulic \cite{ref:LeBo06}. Stationary or rotating point constraints are computed by solving linear equation systems assuming rigid transformations and a uniform noise pattern, itself estimated from known user-defined constraint(s). While addressing the unreliability of common heuristics-based methods due to zero velocity assumption invalidated by noise, this approach is not fully automatic. Moreover, the noise pattern is assumed to be uniform along time, which is not always true.

    Researchers also investigated learned contact detection models. \Eg Ikemoto~\etal~\cite{ref:IkAF06} proposed a K-Nearest Neighbors (KNN) classifier to determine when the feet should be planted and claimed to be more accurate than heuristics-based approaches. However, a very low amount of manually annotated motion data was used (about 3 minutes, single subject) preventing the approach to generalise well. The proposed KNN classifier achieves $90.78\%$ accuracy over about $33$ seconds of hand-labelled motion data, compared to $57.45\%$ and $57.00\%$ for speed-based and height-based thresholding baselines, respectively. However, these relatively low accuracies suggest that the corresponding thresholds are not optimal.

    More recently, researchers leveraged neural networks for exploring this problem. Smith~\etal \cite{ref:SCNW19} detected foot contact labels using a dedicated Multilayer Perceptron (MLP) to remove footskate artefacts through IK in a motion style transfer pipeline. Zou~\etal~\cite{ref:ZYCZ20}, Shi~\etal~\cite{ref:SAAK20} and Rempe~\etal~\cite{ref:RGHR20} concurrently proposed to leverage foot contact detection to refine human motion estimation. In these works, foot contact labels are detected from \dd keypoints, themselves estimated from images using OpenPose \cite{ref:CHSW19}, a state-of-the-art \dd pose estimator. Both Zou~\etal~\cite{ref:ZYCZ20} and Rempe~\etal~\cite{ref:RGHR20} used a dedicated module for contact detection while Shi~\etal~\cite{ref:SAAK20} detected them together with \ddd joint rotations and global root positions. Shimada~\etal~\cite{ref:SGXT20} extended the contact detection module from Zou~\etal~\cite{ref:ZYCZ20} to additionally detect at each frame whether the subject is stationary or not. However, in all these works the foot contact detection modules were trained with ground truth contacts obtained from simple heuristics, as previously described. Tedious manual screening has occasionally been used to correct or label motion sequences, but this approach limits the amount of labelled data which is crucial with neural networks.
    
    Other works also embedded foot contact detection into their model with the goal of improving foot positioning consistency, but do not support detection from motion data. Yang~\etal~\cite{ref:YaKL21} estimated lower-body pose and foot contacts from upper-body AR/VR tracking devices. Harvey~\etal~\cite{ref:HYNP20} proposed a motion in-betweening method where pose and contact labels are interpolated between their known values at user-specified keyframes. Likewise, Holden~\etal~\cite{ref:HoKS17} and Starke~\etal~\cite{ref:SZKS19} include foot contact detection at the next frame in character control frameworks, which enables foot contact continuation. Min and Chain~\cite{ref:MiCh12} modelled foot contacts into a graph-based framework called \emph{Motion Graphs++}, by annotating individual motion primitives with embedded contact information, able to randomly synthesise motion along with contacts at runtime.

\subsection{Ground Reaction Forces Estimation}
\label{section:RelatedWork:GRFs}
    In physics, the force exerted by the ground on a body in contact, such as the human body, is called a ground reaction force (GRF). It is generally difficult to measure but nonetheless important in many fields of study including biomechanics, biomedical engineering and physics-based animation. Researchers in biomechanics and biomedical investigated GRF estimation from plantar pressure sensors \cite{ref:RFCA10, ref:JJLK14, ref:MMPD21}, inertial and optical motion capture systems \cite{ref:FCKD15, ref:KBSd16, ref:FHSC16, ref:MKDB20}, \ddd accelerometers \cite{ref:LBMN15}, and Kinect \cite{ref:EKAO17}. However, GRF estimation for biomechanics or biomedical applications is beyond the scope of our approach that uses vertical GRF (vGRF) distribution as a proxy representation and is intended for human animation applications. For more details on GRF estimation in biomechanics and biomedical engineering, we refer the reader to the systematic review by Ancilloa~\etal~\cite{ref:ATBO18}.
    
    Early works in motion reconstruction leveraged pressure sensors to measure GRFs, because of their importance in dynamics.
    Ha~\etal~\cite{ref:HaBL11} formulated the problem as a per-frame optimisation of end-effector positions obtained from a hand tracking device, and linear and angular momentums measured with pressure platforms.
    Later, Zhang~\etal~\cite{ref:ZSZL14} leveraged a pair of pressure sensing shoes as well as three depth cameras to develop a full-body motion reconstruction framework consisting in kinematic pose reconstruction followed by physics-based motion optimisation.
    
    More recently, several approaches instead estimated GRFs from monocular images, starting with \dd and \ddd pose estimation and then solving physical optimisation problems.
    Zell~\etal~\cite{ref:ZeWR17} proposed to estimate inner and exterior forces by optimising camera parameters and \dd pose reconstruction with a linear combination of base poses in a first step, and then GRFs and inner joint torques to satisfy the equations of motion and resolve camera projection ambiguities.
    Li~\etal~\cite{ref:LSCL19} estimated \ddd motion and forces between a subject and its environment by minimizing the discrepancy between the observed and reprojected \dd poses, with priors on estimated \ddd poses, trajectory smoothness and physical plausibility for regularization.
    Rempe~\etal~\cite{ref:RGHR20} and Shimada~\etal~\cite{ref:SGXT20} proposed a similar pipeline but focusing on more dynamic and diverse human motions, without object interactions. Shimada \etal further correct imbalanced stationary poses.
    Later on, they extended their approach with additional neural networks~\cite{ref:SGXP21}: \emph{TPNet} first regresses target \ddd poses and contact states from \dd keypoints. Then, \emph{GRFNet} and \emph{DyNet} iteratively estimate GRFs and PD controller gain parameters in a dynamic cycle where the character pose is updated at each step after forward kinematics.
    
    Different from motion reconstruction from images, Zell \etal \cite{ref:ZeRW20} proposed a weakly-supervised approach to inverse dynamics. An MLP is trained to estimate GRFs, moments and joint torques from motion such that the input motion is reconstructed using forward dynamics in an optimisation loop. Motion capture data synchronized with force plates enable supervision during training: reconstructed GRF+M and joint torque divergences are penalized while GRF are minimized whenever feet are not in contact with the ground.
    
    To the best of our knowledge, the closest work to the proposed method is the deep learning approach to improve human pose estimation computing stability proposed by Scott \etal \cite{ref:SRFC20}. In this work, body dynamics analysis from joint positions is investigated while most papers on human pose estimation only focus on skeleton kinematics. A convolutional neural network called PressNet is proposed to estimate 2D foot pressure maps from joint positions and validated on a novel dataset of Tai Chi sequences (see Section \ref{section:RelatedWork:Databases}). Center of Pressure (CoP) and Base of Support (BoS) are computed from pressure values and used for validation. Unlike Scott \etal, our work specifically targets foot contact detection and footskate cleanup, tasks that are relevant to human character animation. We explore more diverse motion sequences including different types of locomotion at different paces performed in different ways (forward, backward, sideways) as well as sequences in non-flat environments such as stair climbing and stepping on solid obstacles.

\subsection{Foot Contact \& Ground Reaction Force Databases} \label{section:RelatedWork:Databases}
    As of today, motion capture data annotated with accurate foot contact information are scarce. Researchers in biomechanics and biomedical engineering have released a few databases of motion capture with GRFs \eg \cite{ref:KuSN14}, however most of them are not suitable for animation purposes as they typically focus on specific aspects of movement, or target stage and symptoms recognition of diseases in pathological subjects.

    To the best of our knowledge, the closest database to the one we propose is PSU-TMM100. Recently released to the computer vision community by Scott \etal \cite{ref:SRFC20}, this dataset provides videos from two views, motion capture markers, body joints and foot pressure recorded with insole sensors. It contains about 7.6 hours of data during which 10 subjects are performing 24-form simplified Tai Chi. Although similar in terms of scale and nature of the captured data, the database we propose has quite different types of motion. While PSU-TMM100 contains specific Tai Chi sequences mostly composed of slow body movements with long and stable foot supports, \UnderPressure provides diversified sequences focused on but not restricted to locomotion at different paces including on non-flat environments (see Table~\ref{tab:data}), \ie more challenging conditions for foot contacts detection.
    
    \section{Database}
        \begin{table}[t]
    \small
    \centering
    \caption{
        Motion sequence categories in \UnderPressure.
    }
    \begin{tabular}{c c c}
        Category & Motion Type & Duration [mn] \\
        \hline
        \multirow{4}{2cm}{\centering Locomotion, forwards}
                                    & slow walking & $43.9$ \\
                                    & normal walking & $42.0$ \\
                                    & fast walking & $42.9$ \\
                                    & running & $30.1$ \\
        \cline{2-3}
        \multirow{4}{2cm}{\centering Locomotion, backward}
                                    & slow walking & $21.1$\\
                                    & normal walking & $21.8$ \\
                                    & fast walking & $20.8$ \\
                                    & running & $15.5$ \\
        \cline{2-3}
        \multirow{4}{2cm}{\centering Locomotion, miscellaneous}
                                    & running sideways & $12.3$ \\
                                    & hopping & $13.9$ \\
                                    & stairs 1 at a time & $18.0$ \\
                                    & stairs 2 at a time & $13.9$ \\
        \cline{2-3}
        \multirow{3}{2cm}{\centering Locomotion with obstacles}
                                    & stepping on obstacles & $5.7$ \\
                                    & stepping over obstacles & $11.4$ \\
                                    & jumping over obstacles & $10.5$ \\
        \cline{2-3}
        \multirow{4}{*}{Idle}
                                    & leg stand-up & $4.8$ \\
                                    & sit-down & $4.8$ \\
                                    & crouched down & $4.8$ \\
        \hline
        Total & & 338.2
        
    \end{tabular}
    \label{tab:data}
\end{table}

\begin{figure}[t]
    \centering
    \begin{subfigure}[b]{0.45\linewidth}
        \centering
        \includegraphics[width=0.95\linewidth]{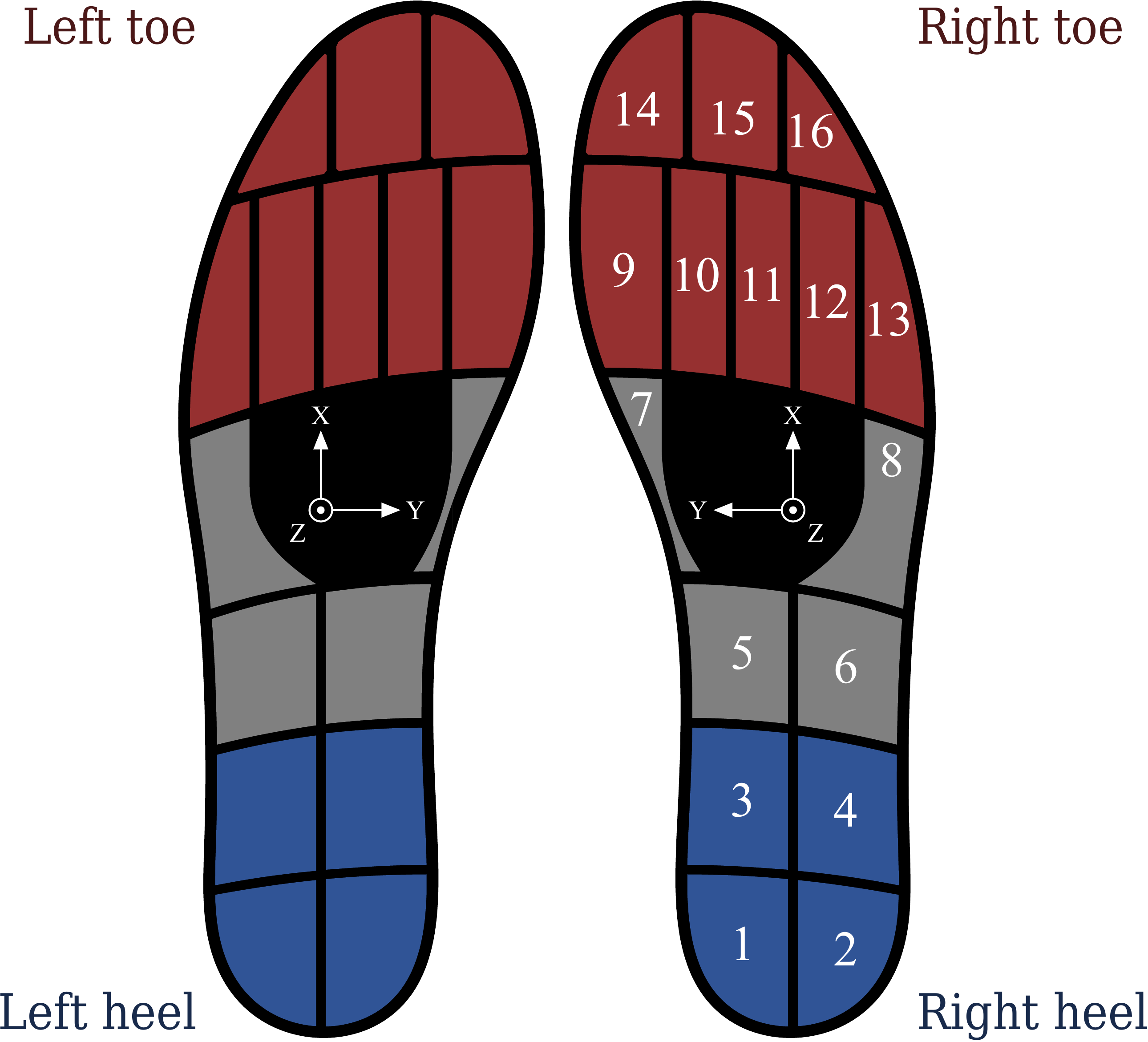}
        \caption{Insoles}
        \label{fig:insoles}
    \end{subfigure}
    \begin{subfigure}[b]{0.54\linewidth}
        \centering
        \includegraphics[width=0.95\linewidth]{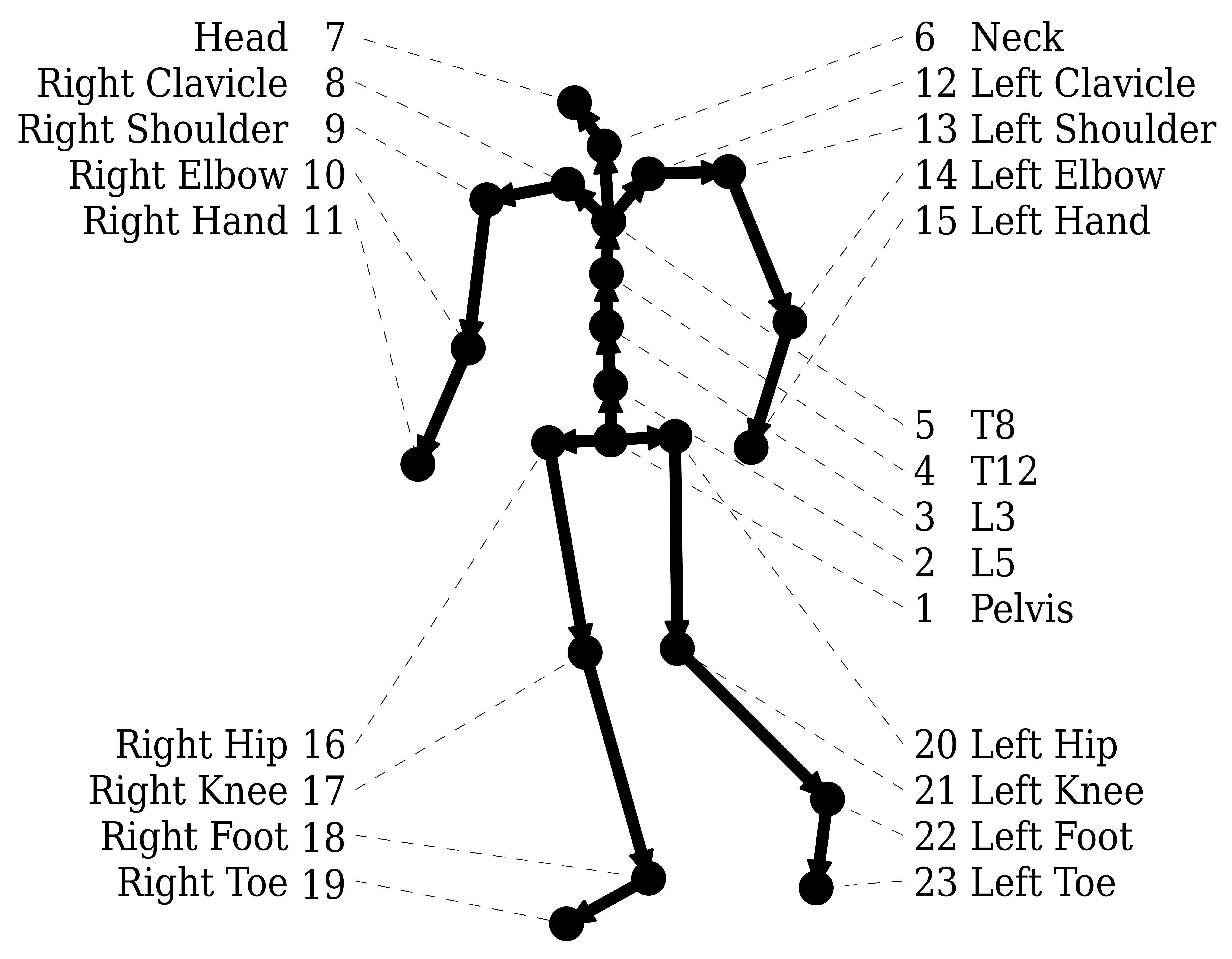}
        \caption{Skeleton}
        \label{fig:skeleton}
    \end{subfigure}
    \caption{
        Left) Pressure cell layout of \emph{Moticon's OpenGo Sensor Insoles} \cite{ref:OpenGo}. Blue (1 to 4) and red (9 to 16) cells are the groups of cells used to compute heel and toe contacts, respectively. Axes at insole centers represent inertial measurement units (IMUs).
        Right) \emph{Xsens MVN}'s \cite{ref:ScGB18} motion capture skeleton with 23 joints.
    }
    \label{fig:data}
\end{figure}

In this work, we release \UnderPressureWithUrl, a motion capture database annotated with pressure insoles data, designed primarily for character animation purposes. In the following, we provide information about the capture, the motion characteristics, and the preprocessing steps.

We recorded $10$ healthy adult volunteers (2F, 8M) with diverse morphologies aged between $21$ and $55$ years ($32 \pm 11~yr$), weighing between $65$ and $91$ kilograms ($79\pm 9~kg$), and measuring between $167$ and $187$ centimeters ($177 \pm 5~cm$). Each subject performed the same set of activities, including forward and backward locomotion at different paces, sitting, standing, passing obstacles, climbing stairs, as well as motions on uneven terrain like going up and down stairs. The detailed composition of our dataset is provided in Table~\ref{tab:data}. Motion capture data for each subject last approximately $34$ minutes, for a total of $5.6$ hours of motion capture.

\begin{figure*}
    \centering
    \includegraphics[width=0.8\linewidth]{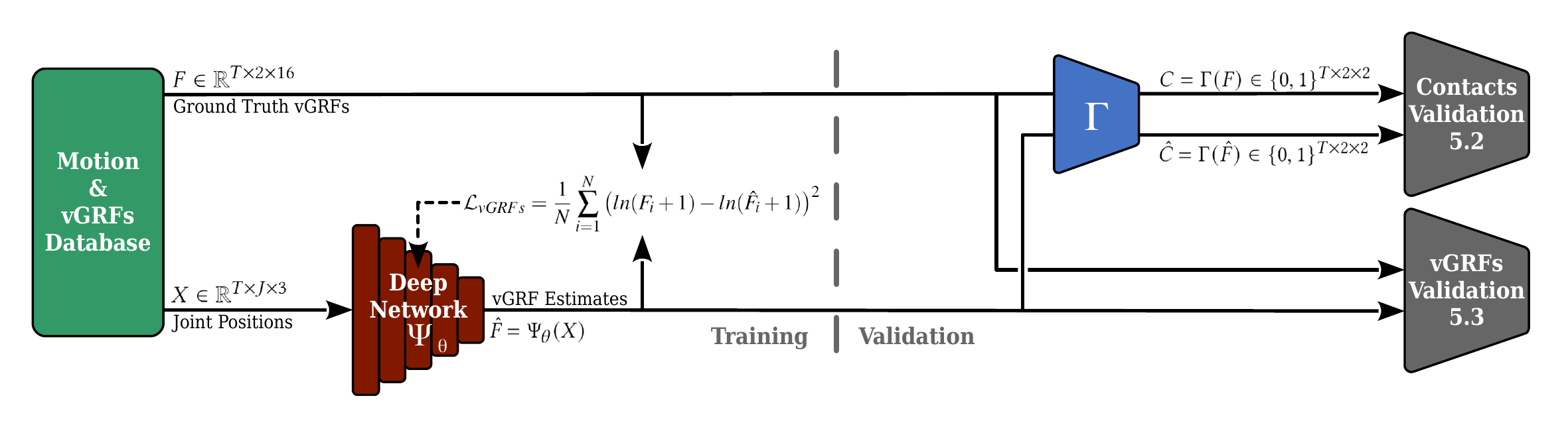}
    \caption{
        Overview of our approach. Synchronized vGRFS and motion capture database (green rectangle) provide inputs and targets to train deep network $\Psi_{\theta}$ depicted in red. The blue trapezoid represents the contact function $\Gamma$ (see Section~\ref{section:Post-CaptureProcessing}). At runtime, our deep network augments motion data with vGRFs from which foot contact labels can be derived using the contact function $\Gamma$, both useful in many applications, \eg, reconstructing motion from images, cleaning footskate, finding suitable transition frames for motion blending, adapting animations to uneven terrain, and many more. As illustrated, both estimated vGRFs and detected contacts are evaluated in Sections~\ref{section:Results:Contacts} and~\ref{section:Results:vGRFs}, respectively.
    }
    \label{fig:overview}
\end{figure*}

\subsection{Motion Capture}
    Subjects were equipped with an \emph{Xsens MVN Link} motion capture system~\cite{ref:ScGB18}. The hardware consists of $17$ inertial measurement units (IMUs) running at $240~Hz$ embedded in the \emph{MVN Link} suit. Each IMU contains a \ddd accelerometer, a gyroscope, and a magnetometer. Capturing was performed using the \emph{Xsens MVN Animate} software, an engine customised for \ddd character animation that combines tracking data of the $17$ individual IMUs with a $23$-segment biomechanical model to obtain segment positions and orientations. We calibrated \emph{MVN Animate} for each subject with height, arm span and shoe length measurements as inputs while other body dimensions and proportions were estimated through the calibration, as well as the orientation of motion trackers w.r.t. the corresponding segments. After \emph{MVN Animate} processing, motion data consist of pose sequences with $23$ segments sampled at $240~Hz$.
    
\subsection{Foot Pressure} \label{section:FootPressure}
    In addition to motion capture, we recorded the spatial distribution of plantar foot pressures. To this end, subjects were also equipped with \emph{Moticon's OpenGo Sensor Insoles} \cite{ref:OpenGo} placed into their shoes. Each insole has $16$ plantar pressure sensors with a resolution of $0.25~N/cm^2$ and a $6$-axis inertial measurement unit (IMU), both running at $100~Hz$ (See Figure~\ref{fig:insoles}). Moreover, we weighed each subject with full equipment to enable vGRFs normalisation and equipped subjects with the same shoes whose soles are thin and flexible insole for accurate and faithful pressure measures as well as for controlling the grip.
    
\subsection{Post-Capture Processing} \label{section:Post-CaptureProcessing}
    \paragraph*{Vertical GRFs}
        Captured data include motion sequences, plantar pressure distribution and foot acceleration. Since pressure is defined as the perpendicular force per unit area, we additionally compute vertical GRF components (vGRFs) by multiplying pressure values by the corresponding cell areas. The motivation here is that groups of vGRFs are easier to aggregate (by summation). We also normalise these values to express them as subject weight proportions.
    
    \paragraph*{Foot Contact Labels}
        We derive ground truth foot contact labels deterministically from vGRFs. In this work, we consider two contact locations per foot, \ie heels and toes as commonly done in human animation \cite{ref:IkAF06, ref:HoKS17, ref:SGXP21}. To compute contact labels, vGRF components are first smoothed with a Gaussian filter to avoid rapidly alternating labels due to threshold effects. Then, smoothed vGRF components are summed per contact location (see red and blue cells in Figure~\ref{fig:insoles}), rescaled such that the sum of blue and red cell vGRFs is equal to total vGRF (to properly ignore gray cells which particularly suffer from noised measures and can be activated during either toes or heels contact) and then a threshold at $5\%$ of the body weight is applied to obtain raw labels. Finally, raw labels are discarded whenever per-foot total vGRF (including gray cells) is below $10\%$ of the body weight to avoid false positives triggered by noise. Contact phases shorter than $0.1s$ are also discarded for the same reason. In the following, we refer to this binary contact labels calculation as the contact function $\Gamma$.
   
    \paragraph*{Synchronization}
        Since we jointly capture motion and foot pressure data with separate devices, our records must be accurately synchronized in absence of a genlock signal. To this end, subjects were asked to perform a simple control movement at the beginning and end of each capture sequence, consisting in an in-place double-leg jump. This enables matching vertical acceleration peaks measured on pressure insole IMUs with peaks computed from motion captured foot positions. Although numerical differentiation is known to amplify high frequency noise, we found that the framerate of our motion capture data was sufficiently high and measurement noise was small enough for synchronization.

    \paragraph*{Downsampling and Trimming}
        After synchronizing our data, we downsampled motion capture data from $240~Hz$ to $100~Hz$ using spherical linear interpolation to match the framerate of pressure insole data. Original motion sequences at $240~Hz$ are also provided in our database. We also automatically trimmed the beginning and end of each sequence to remove the synchronization patterns.
    \section{Deep Neural Network}
        In this section, we describe the proposed method to learn a deep neural network model estimating vGRF distributions from motion capture data. Learning vGRFs instead of binary contact labels encourages our deep neural network to more accurately model interactions between feet and ground, and enforces motion dynamics understanding. See Figure~\ref{fig:overview} for an overview of our approach.

\subsection{Data Representation}
    \paragraph*{Input}
        At each frame $t$, the human pose $X_{t} \in \mathbb{R}^{J \times 3}$ is represented by the position of its $J=23$ joints in a global Euclidean space. We design our deep network $\Psi_{\theta}$ to output vGRFs and contact labels at each frame from a few surrounding input frames with padding when needed. The full input pose sequence is then $X \in \mathbb{R}^{T \times J \times 3}$, where $T$ is a variable number of frames.
        
    \paragraph*{Output}
        As previously described, our deep network $\Psi_{\theta}$ estimates the vGRF distribution from motion data. At each frame, it outputs $\hat{F} = \Psi_{\theta} \left( X \right)$ with $\hat{F} \in \mathbb{R}^{T \times 2 \times 16}$, \ie $16$ positive real-valued vGRF components, corresponding to the $16$ insole pressure cells for each foot, expressed proportionally to subject weight.
        
\subsection{Network Architecture}
\label{section:Method:Architecture}
    We designed our network $\Psi_{\theta}$ to process variable-length sequences. To this end, the network is composed of four 1D temporal convolutional layers with $7$-frame wide kernels, followed by three fully-connected layers applied at each frame independently to preserve the support variable-length sequence. Each convolutional or fully-connected layer is followed by exponential linear units (ELU) as activation, except for the last one which is a softplus activation to output nonnegative vGRF components.

\subsection{Training and Inference}
\label{section:Method:TrainingAndInference}
    During training, our network is iteratively exposed to sequences of human poses and tries to estimate corresponding vGRF components as depicted in Figure~\ref{fig:overview}. To encourage robustness and smooth convergence, we make use of stochastic data augmentation. First, similar to random crops and rotations used on images in computer vision, we apply random vGRF-invariant transformations on input pose sequences including translations, horizontal rotations, scaling, and left-right mirroring.

    For each sequence, we also randomly draw its skeleton which is then animated by joint angles to further robustify our network. To do so, we precomputed (offline) an SVD basis of the skeletons captured in our database. At training time, we draw new skeletons by linearly combining precomputed singular vectors with randomly sampled weights. We then further edit these skeletons by randomly moving joint relative positions and rescaling bone lengths to obtain morphology variations. The resulting input motion sequences purposely suffer from artefacts since kinematic chains (\ie from root joint to feet) have been randomly edited, which encourages the network to be resilient w.r.t. perturbed inputs. Joint positions are finally computed through forward kinematics and fed to the network. 
    
    To train our deep network $\Psi_{\theta}$, we minimize a reconstruction loss of vGRF components. Instead of the standard mean squared error (MSE), we minimize the mean squared logarithmic error (MSLE). It has the property to only focus on the relative difference between target and estimated values (see right-hand side of Equation~\ref{eq:loss}), which is convenient when the values considered can be several orders of magnitude apart. In our case, actual vGRFs can be strictly positive and arbitrarily low (\eg during transition from the double leg stance to the single leg stance) as well as very high (\eg during jump landing). The loss function used to train our network is then
    \begin{equation} \label{eq:loss}
        \mathcal{L} = \frac{1}{N}\sum_{i=1}^{N}{ \left( ln(F_{i}+1) - ln(\hat{F}_{i}+1) \right) }^2 = \frac{1}{N}\sum_{i=1}^{N}{ ln ^2 \left( \frac{F_{i}+1}{\hat{F}_{i}+1} \right)}
    \end{equation}
    where $F_i$ are the ground truth vGRF components, and $\hat{F}_i = \Psi_{\theta}(X)_i$ their estimated counterpart. Adding 1 to both $F$ and $\hat{F}$ ensures that the loss is defined when vGRF component value goes to zero.

    At inference, our deep network estimates vGRF components from joint positions as inputs. Then, foot contact labels can be calculated from vGRF estimates using the contact function $\Gamma$ (see Section \ref{section:Post-CaptureProcessing}).
    \section{Evaluation}
        In this section we present results of our method to assess estimated vGRFs and detected foot contacts. After providing implementation details necessary for reproducibility in Section~\ref{section:ImplementationDetails}, we assess foot contact labels detection in Section~\ref{section:Results:Contacts} then vGRFs estimation in Section~\ref{section:Results:vGRFs} on ground truth motion sequences. In Section \ref{section:Results:Perturbations}, we also evaluate foot contact detection performance on different perturbed motion sequences, simulating challenging conditions encountered in concrete applications. Qualitative results illustrating estimated vGRFs and foot contact labels as well as sample motion sequences of our database are available in the supplementary material.

\subsection{Implementation Details}
\label{section:ImplementationDetails}
    \paragraph*{Data}
        We split the proposed \UnderPressure database into two subsets, \ie a training set and a testing set. To ensure robust evaluation, the testing set is composed of the sequences performed by three out of the ten subjects (1F+2M, $\{ S8, S9, S10\}$), representing approximately $30\%$ of the overall database. For training, we further divide the remaining $70\%$ to keep a validation set ($\sim 10\%$) and use early stopping during training. Moreover, we split each training motion sequence into overlapping windows of $T=240$ frames, \ie $2.4 s$.
        
    \paragraph*{Architecture}
        The four convolutional layers at the beginning of our network have respectively 128, 128, 256, and 256 $7$-frame wide filters while the back-end fully-connected layers have 256 neurons each. Dropout with probability $p=0.2$ is applied before each fully-connected layer. The total number of weights in our network is approximately $1.1$ million.
    
    \paragraph*{Training}
        We implemented our deep neural network using PyTorch. Training and validation were executed on an NVidia Tesla V100 GPU while other results were obtained either on an NVidia GeForce RTX 2060 GPU or on CPU. We trained our deep neural network through stochastic gradient descent (see Equation~\ref{eq:loss}) for about $2500$ epochs (about $10$ days) at each of which a new version of the training set was randomly generated (see Section \ref{section:Method:TrainingAndInference}). We used Adam optimization with a batch size of $64$, learning rate $\alpha = 3*10^{-5}$ and hyperparameters $\beta_1 = 0.9$ and $\beta_2 = 0.999$.

\subsection{Foot Contacts Detection}
\label{section:Results:Contacts}
    
    \paragraph*{Metric}
        To evaluate foot contact detection, we use the $F_1$ score which is the harmonic mean of precision (fraction of correctly detected labels among detected labels) and recall (fraction of correctly detected labels among expected contact labels).
    
    \paragraph*{Baseline}
        To evaluate our model against commonly used heuristics-based approaches using thresholds (see Section \ref{section:RelatedWork}), we define an \emph{optimal thresholds (OT)} baseline. This model has two parameters that are thresholds on foot height relative to the ground and velocity norm. To demonstrate the effectiveness of our approach, we set these two parameters to the values which maximize the $F_1$ score over the training set, computed by recursive grid search. Note that in practical applications, such optimal thresholds are not available and must be guessed (hence our naming heuristics-based). This OT baseline therefore constitutes an upper bound in terms of performance over the $F_1$ score of such threshold-based approaches.
    
    We also investigate the relevance of the proposed architecture with an ablative study. We first consider two linear models taking as inputs joint positions and velocities, as learned generalisations of the OT baseline: the former, called \emph{Linear-Feet}, takes only foot and ankle joints as inputs while the latter, called \emph{Linear}, takes all joints. Then, we consider a \emph{3-layer MLP} model, \ie the architecture of the foot contact label detection module proposed by Smith \etal \shortcite{ref:SCNW19} in their style transfer framework, which has three 128-neurons-wide layers and ReLU activations, introducing non-linearities w.r.t. Linear-Feet and Linear models. These three models have real-valued outputs, for which positive values are considered as foot contacts with the ground, and are trained with a binary cross-entropy (BCE) loss. Finally, we explore two variants of our deep network. First, the \emph{Ours-C} variant has the architecture proposed in Section \ref{section:Method:Architecture} except for the last layer and activation which are adapted (\ie number of neurons and sigmoid instead of softplus) to detect foot contact labels instead of estimating vGRF components, and is trained with BCE loss. The second variant corresponds to the combination of \emph{Ours-C} with the main proposed model \emph{Ours}. In this extended variant called \emph{Ours-C\&F}, our architecture is adapted to output both contacts and vGRF components (convolutional layers are shared while fully-connected layers are duplicated and separately learned for each output) and trained with both MSLE (see Section \ref{section:Method:Architecture}) over vGRF components and BCE over contact labels.
    
    \newcommand{\intercol}{\hskip 0.20cm}
\begin{table}
    \small
    \centering
    \caption{
        $F_1$ score on foot contact labels detection of our method, its variants for ablative study purposes, and the OT baseline. Bold and underline respectively indicate per-column best and second best. Our method outperforms the OT baseline and its linear generalisations, and the proposed architecture seems relevant w.r.t. the 3-layer MLP.
    }

    \begin{tabular}{c | c @{\intercol} c @{\intercol} c @{\intercol} c @{\intercol} c @{\intercol} c | c}

        Model & \rotatebox{90}{Walking} & \rotatebox{90}{Running} & \rotatebox{90}{Obstacles} & \rotatebox{90}{Hopping} & \rotatebox{90}{Stairs} & \rotatebox{90}{Idle} & \rotatebox{90}{Overall} \\
        \hline

    	OT baseline & 0.927 & 0.869 & 0.926 & 0.859 & 0.882 & 0.826 & 0.909 \\			
    	\hline		
    	Linear-Feet & 0.936 & 0.863 & 0.921 & 0.824 & 0.906 & 0.812 & 0.913 \\
    	Linear & 0.937 & 0.868 & 0.926 & 0.855 & 0.925 & 0.912 & 0.923 \\
    	3-layer MLP & 0.940 & 0.883 & 0.926 & 0.882 & 0.947 & 0.921 & 0.930 \\		
    	\hline
    	Ours-C & 0.946 & 0.917 & 0.941 & \underline{0.930} & \underline{0.956} & 0.941 & 0.942 \\
    	Ours-C\&F & \underline{0.948} & \underline{0.918} & \underline{0.942} & 0.923 & 0.954 & \underline{0.946} & \underline{0.943} \\
    	Ours & \textbf{0.949} & \textbf{0.930} & \textbf{0.944} & \textbf{0.931} & \textbf{0.959} & \textbf{0.948} & \textbf{0.947}

    \end{tabular}
    \label{tab:contacts}
\end{table}





    
    Table~\ref{tab:contacts} reports the $F_1$ score for each model and each motion category as well as overall results (right-most column). First, learned linear regressions show improved performances w.r.t. the OT baseline, which tends to confirm that thresholds based approaches lack complexity. The linear model (all joints) improvement w.r.t. Linear-Feet (only foot and ankle) suggests that contact labels detection can also benefit from other body joints, pointing out one of the limitation of thresholds based approaches that typically process only foot joints. The higher $F_1$ score of the 3-layer MLP model w.r.t. linear models confirms their limitations for foot contact label detection. Finally the proposed architecture further increases the detection accuracy with relatively small differences among variants.
    
    \begin{figure}
    \centering
    \scriptsize
    \begin{tikzpicture}
        \node (graphic) {\includegraphics[width=0.97\linewidth]{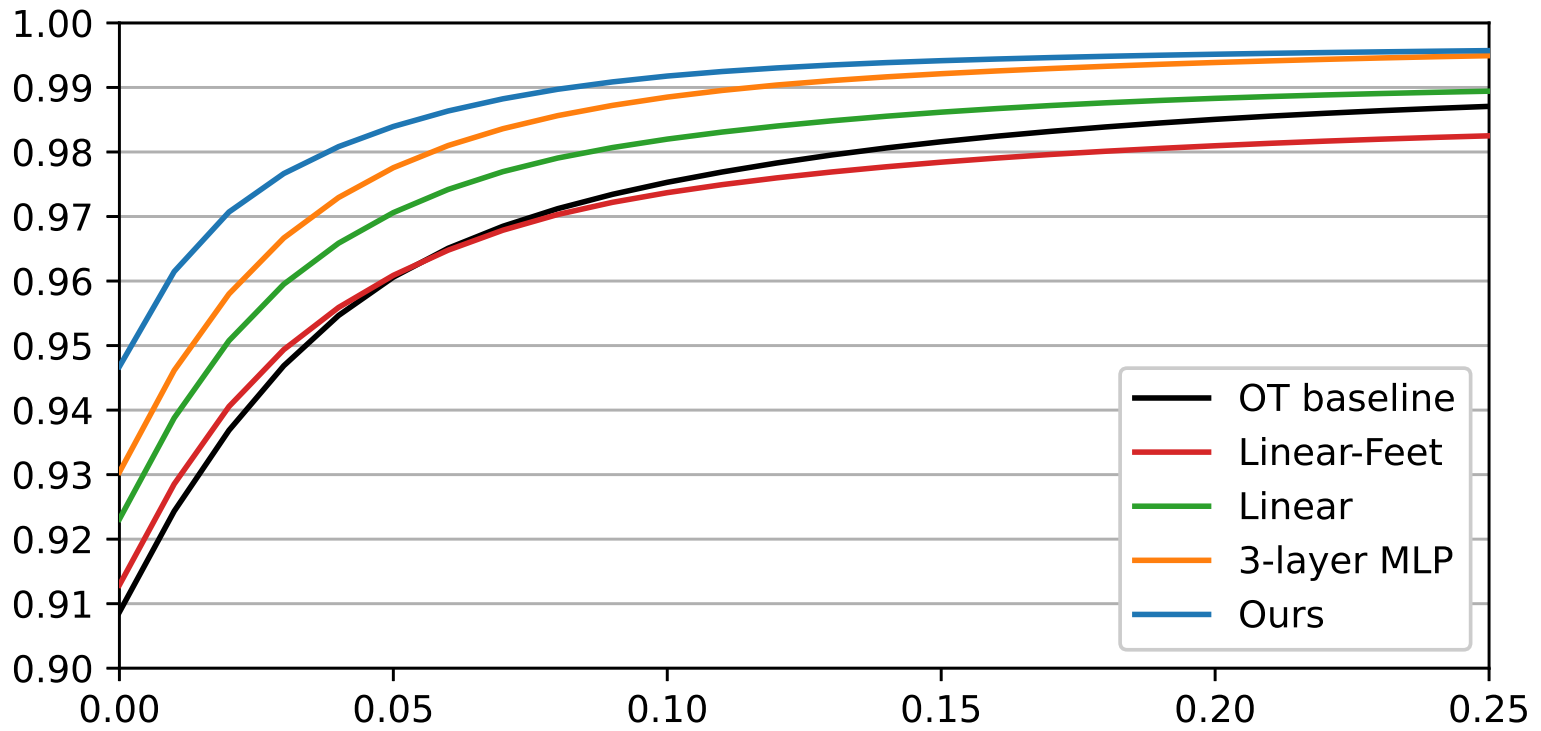}};
        \node[below=of graphic, node distance=0cm, yshift=1.1cm] {Tolerance [s]};
        \node[left=of graphic, node distance=0cm, rotate=90, anchor=center,yshift=-0.95cm] {F$_1$ score};
    \end{tikzpicture}
    \caption{
        $F_1$ score curves against temporal tolerance. At any given tolerance $t$, the $F_1$ score is computed with contact labels considered as incorrect if and only if they are wrong and located at least $t$ seconds away from the closest contact phase. Without any tolerance (left-most), $F_1$ scores correspond to the \emph{Overall} column in table \ref{tab:contacts}. The OT baseline and its linear generalisations have large errors far from contact phase changes, resulting in relatively low $F_1$ scores compared to our model even with high temporal tolerances.
    }
    \label{fig:F1_curves}
\end{figure}
    \begin{figure}
    \centering
    \scriptsize
    \begin{tikzpicture}
        \node (graphic) {\includegraphics[width=0.97\linewidth]{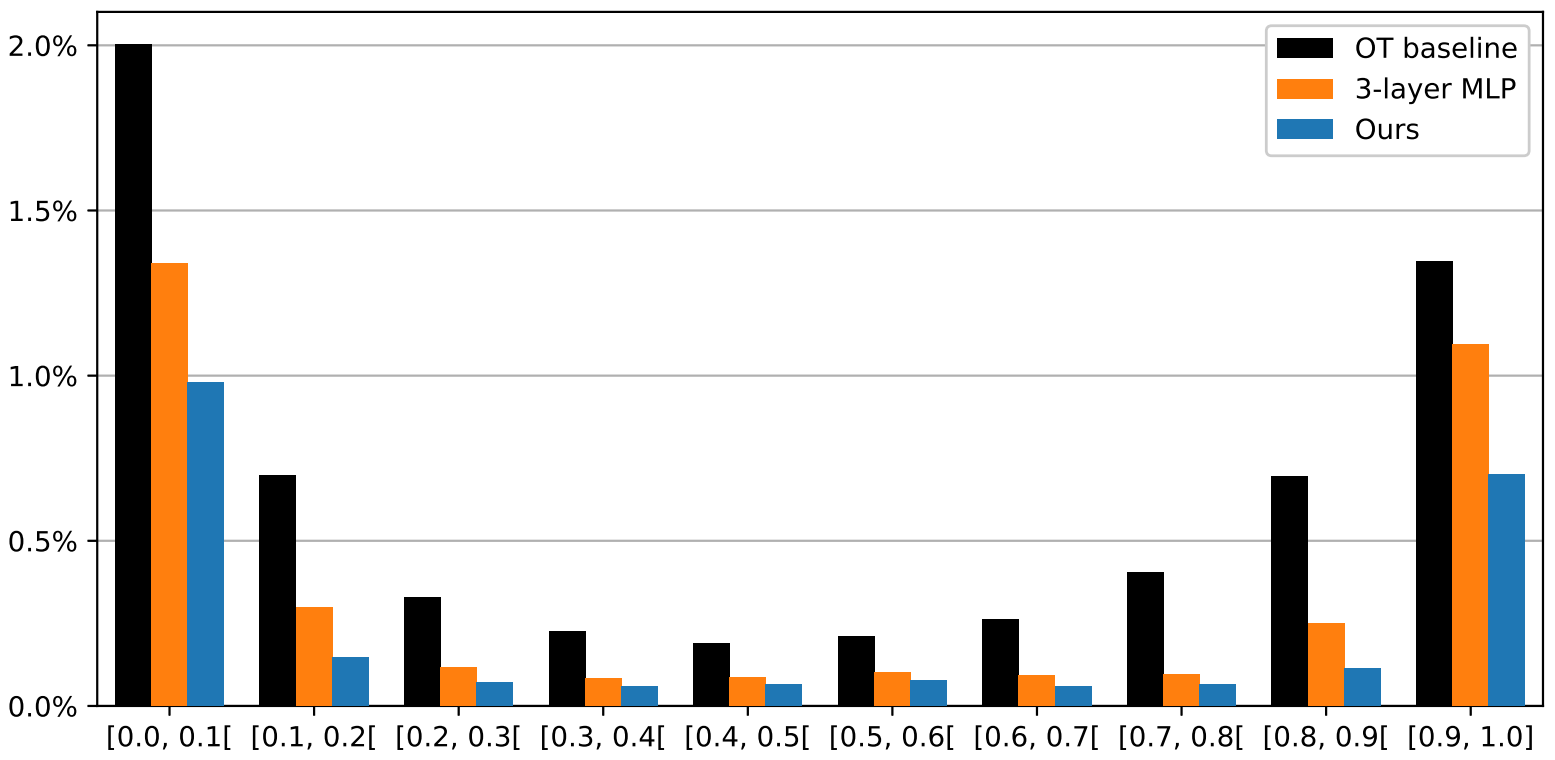}};
        \node[below=of graphic, node distance=0cm, yshift=1.1cm] {Normalised time};
        \node[left=of graphic, node distance=0cm, rotate=90, anchor=center,yshift=-0.95cm] {False positive rate};
    \end{tikzpicture}
    \caption{
        False positive rate distributions w.r.t. normalised off-contact phases. Ground truth off-contact phase intervals are mapped to $[0, 1]$ and false positives are aggregated in this normalised temporal frame. Intuitively, the farther false positives are from the closest contact phase (\ie $0$ or $1$ in the normalised temporal frame), the more severe they are. False positive rates of learned models like ours quickly decrease while the OT baseline keeps it significantly higher in-between contacts.
    }
    \label{fig:error_distr}
\end{figure}

\paragraph*{Temporal Analysis of Foot Contact Detection}
   Up to now, we provided temporally global foot contact detection results. However, misdetected labels located closer to contact phase changes are less severe in the sense that they would result in less severe biases or artefacts in most applications, \eg footskate cleanup. To this end, we provide a finer analysis of foot contact labels. In Figure~\ref{fig:F1_curves}, we plot the $F_1$ score against an increasing temporal tolerance to detection errors. In Figure~\ref{fig:error_distr}, the distribution of false positive rates is given according to a normalised temporal frame where ground truth off-contact phase intervals are mapped to $[0, 1]$. Both figures confirm the limitations of heuristics approaches represented by the OT baseline whose false positive rate is significant in the middle of off-contact phases. On the contrary, learned models show convergence of accuracy close to $100\%$ with increasing tolerance as well as low false positive rates in the middle of off-contact phases.

\renewcommand{\intercol}{\hskip 0.15cm}
\begin{table}
    \small
    \centering
    \caption{
        Root mean squared error of the estimated vGRF normalised by body weight. 
        Estimating foot contacts in addition to vGRFs (Ours-C\&F) seems slightly detrimental compared to estimating only vGRFs (Ours).
    }

    \begin{tabular}{c | c @{\intercol} c @{\intercol} c @{\intercol} c @{\intercol} c @{\intercol} c | c}

        Model & \rotatebox{90}{Walking} & \rotatebox{90}{Running} & \rotatebox{90}{Obstacles} & \rotatebox{90}{Hopping} & \rotatebox{90}{Stairs} & \rotatebox{90}{Idle} & \rotatebox{90}{Overall} \\
        \hline

    	Ours-C\&F & 9.5\% & 14.8\% & 12.4\% & 14.3\% & 11.8\% & 13.1\% & 11.4\% \\
    	Ours & \textbf{9.1\%} & \textbf{14.3\%} & \textbf{11.6\%} & \textbf{14.1\%} & \textbf{10.9\%} & \textbf{11.9\%} & \textbf{10.9\%}

    \end{tabular}
    \label{tab:vGRF}
\end{table}

\subsection{vGRFs Estimation}
\label{section:Results:vGRFs}
   \paragraph*{Metrics}
        In this section we assess performances of our method on vGRFs estimation from motion capture data. We use the root mean squared error (RMSE) of per-foot total vGRF which is mainly sensitive to global biases (at foot scale). In complement, we also evaluated the Center of Pressure (CoP) computed from estimated vGRF components, which is mostly sensitive to local errors (at pressure cell scale) but we provide the corresponding results in the supplementary material.

    Table~\ref{tab:vGRF} gives the RMSE of the estimated vGRFs proportionally to the subjects' body weight. As for contact detection, these results suggest that learning to model foot contact labels in addition to vGRF (Ours-C\&F) reduces accuracy, although the performances of both variants are close. Note that other variants evaluated on contact detection in the previous section only detect contact labels and hence cannot be evaluated on vGRF estimation.
    
    In biomechanics, force plates are considered gold standard to measure vGRF and CoP; however, the environment in which we captured the proposed database (significant area including obstacles and stairs) prevented us to use force plates. For this reason, our evaluation considers pressure insole measures as ground truth for evaluation purposes. Existing works in biomechanics \cite{ref:NaSM11, ref:JMNB19} evaluated pressure insoles accuracy w.r.t. force plates, and tell us that vGRFs measured with pressure insoles suffer from a RMSE up to approximately $10\%$ of the subject weight, being subject to variations depending on experimental conditions. Thus, Table~\ref{tab:vGRF} indicate that vGRF estimation errors of our deep neural network are approximately of the same order of magnitude as the measurement error expected with pressure insoles.

\subsection{Foot Contacts Detection in Challenging Conditions}
\label{section:Results:Perturbations}
    As explained in Section \ref{section:RelatedWork}, many applications requiring foot contact labels detection consist in quantifying or correcting foot artefacts. By definition, motion sequences to be processed in such applications are not expected to be flawless. As a consequence, contact detection performance is not solely relevant for clean motion capture data (as evaluated in Section \ref{section:Results:Contacts}), but also for perturbed sequences. In this section, we evaluate the performance of our model on motion sequences in which we introduce three types of artefacts: additive Gaussian noise, distorsions caused by going through a partially trained autoencoder, and artefacts (mostly footskate) obtained by blending different motion sequences.
  
\begin{figure}
    \centering
    \scriptsize
    \begin{tikzpicture}
        \node (graphic) {\includegraphics[width=0.96\linewidth]{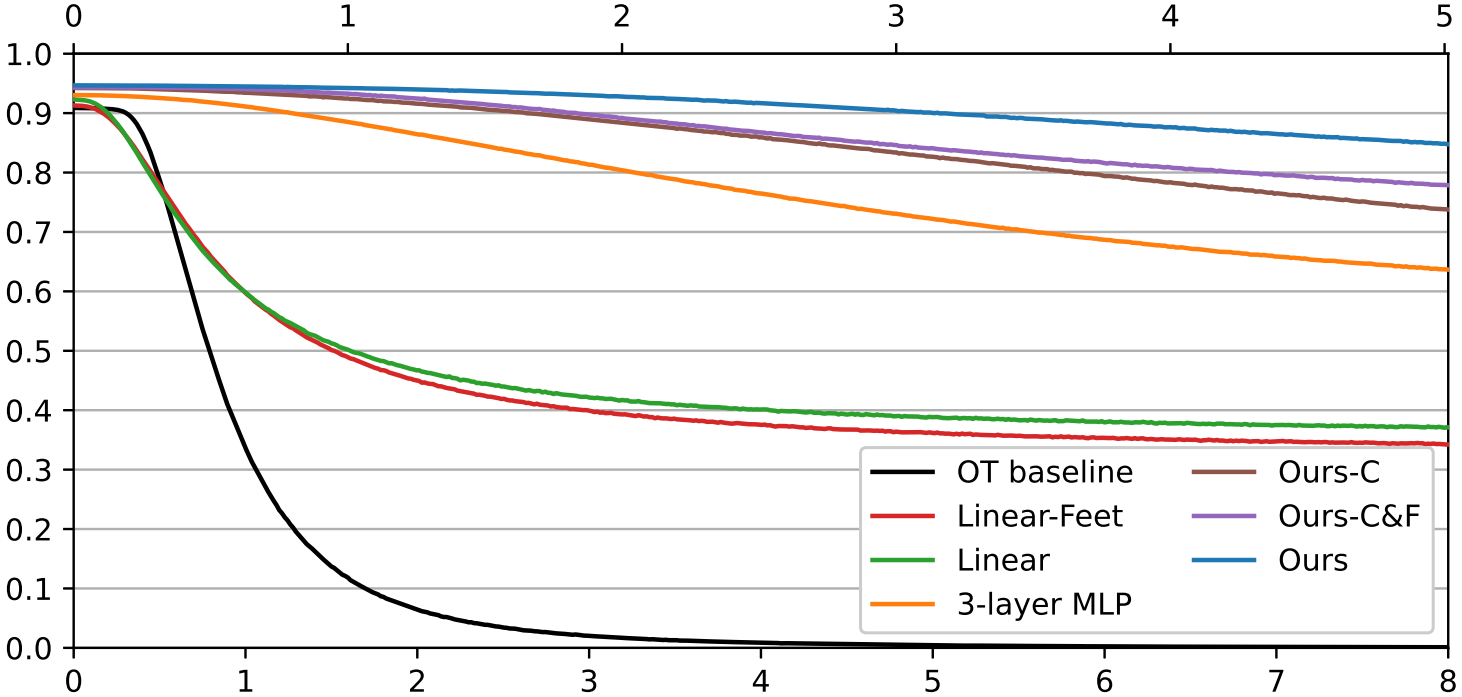}};
        \node[below=of graphic, node distance=0cm, yshift=1.1cm] {MPJPE [cm]};
        \node[left=of graphic, node distance=0cm, rotate=90, anchor=center,yshift=-0.95cm] {F$_1$ score};
        \node[above=of graphic, node distance=0cm, yshift=-1.1cm] {Standard deviation [cm]};
    \end{tikzpicture}
    \caption{
        $F_1$ score on foot contacts detection from motion sequences purposely noised with additive isotropic Gaussian noise. Each curve represents the $F_1$ score against the amount of noise introduced, measured with the MPJPE in centimeters indicated by the bottom horizontal axis, while the top horizontal axis gives the corresponding standard deviations of the Gaussian noise. The results indicate that our method is more robust to noise.
    }
    \label{fig:noised}
\end{figure}

\begin{figure}
    \centering
    \scriptsize
    \begin{tikzpicture}
        \node (graphic) {\includegraphics[width=0.96\linewidth]{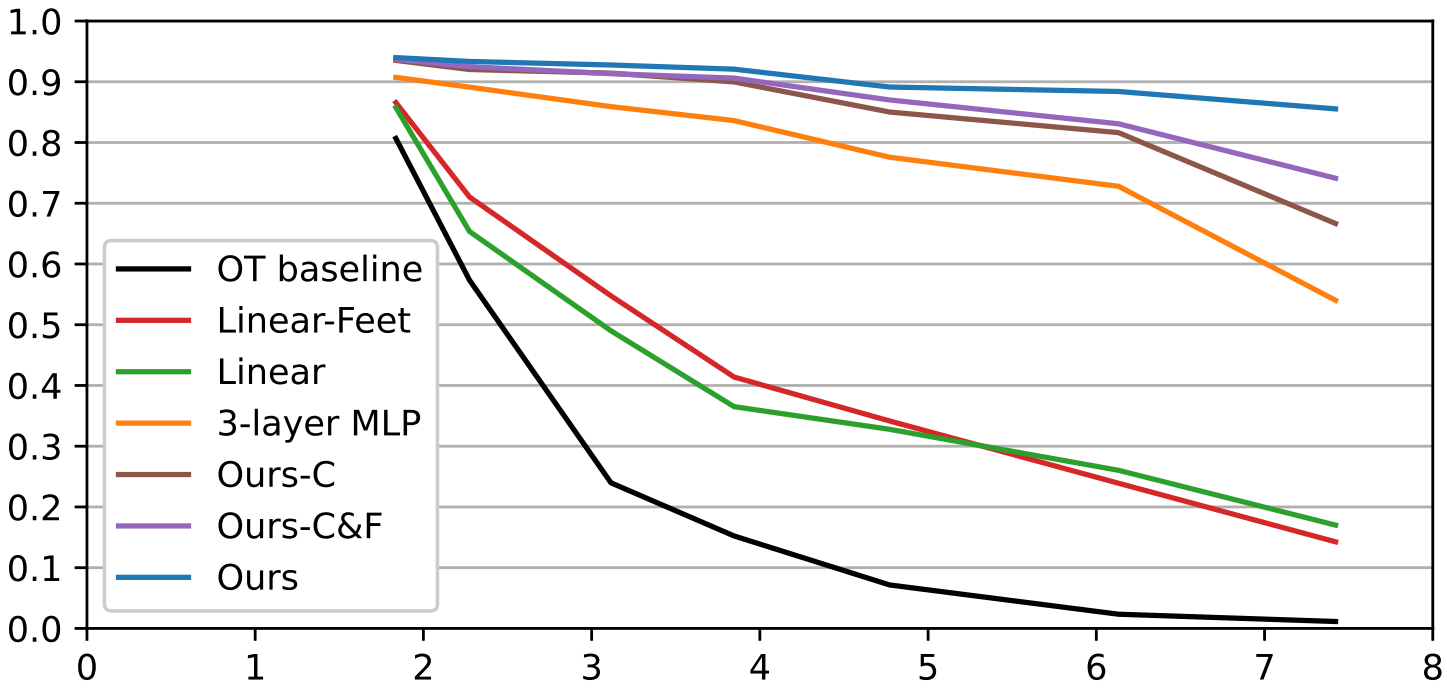}};
        \node[below=of graphic, node distance=0cm, yshift=1.1cm] {MPJPE [cm]};
        \node[left=of graphic, node distance=0cm, rotate=90, anchor=center,yshift=-0.95cm] {F$_1$ score};
    \end{tikzpicture}
    
    \caption{
        $F_1$ score on foot contacts detection from motion sequences distorted by a motion autoencoder early-stopped at different epochs to emulate different amounts of distorsion. Each curve displays the $F_1$ score against the amount of distorsion introduced, measured with the MPJPE in centimeters. The results indicate that our method is more robust to distorted sequences.
    }
    \label{fig:autoencoder}
\end{figure}

    \paragraph*{Motions Perturbed with Gaussian Noise}

    First, we simply evaluate how the contact detection performance of the various models evaluated in Section~\ref{section:Results:Contacts} is affected by an increasing amount of Gaussian noise added to joint positions. Figure~\ref{fig:noised} displays the $F_1$ score obtained by the different models against the amount of noise introduced.
    
    \paragraph*{Motions Distorted by an Autoencoder}
        To simulate perturbations more faithful to what is encountered in real-world applications, we trained an autoencoder to reconstruct motions from our testset. We used a similar architecture to the convolutional autoencoder proposed by Holden~\etal~\cite{ref:HoSK16} and trained it for 100 epochs (about 10 minutes). We kept a selection of epochs at which the amount of perturbations introduced by the autoencoder follows a nice monotonically decreasing curve to avoid outlier epochs. At each of these epochs, we evaluate foot contact detection on motion sequences that have been encoded and then decoded to introduce increasing perturbations. Again we use the $F_1$ score as a foot contact labels detection performance metric and display in Figure~\ref{fig:autoencoder} the results against the observed amount of noise introduced by the autoencoder at different epochs.

\begin{figure}
    \centering
    \scriptsize
    \begin{tikzpicture}
        \node (graphic) {\includegraphics[width=0.96\linewidth]{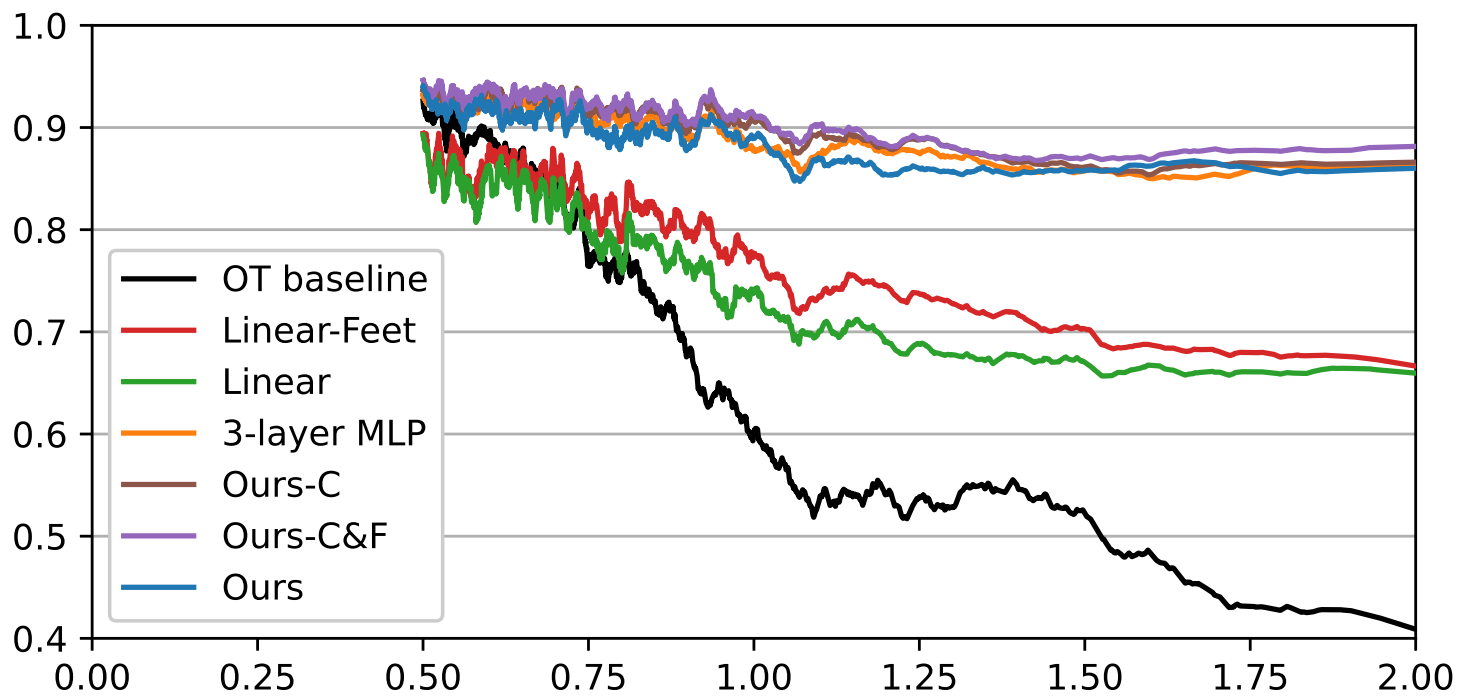}};
        \node[below=of graphic, node distance=0cm, yshift=1.1cm] {Mean velocity of the feet during contact phases [m/s]};
        \node[left=of graphic, node distance=0cm, rotate=90, anchor=center,yshift=-0.95cm] {F$_1$ score};
    \end{tikzpicture}
    
    \caption{
        $F_1$ score against amount of footskate. Test motions with matching foot contact patterns have been blended to purposely introduce footskate while preserving contact labels. Foot contact detection is then evaluated on such motions suffering from footskate, and compared to the ground truth contact labels. The amount of footskate introduced through blending is measure using the mean velocity of the feet during contact phases. Non-linear learned detection models keep reasonably high accuracy while accuracy of linear models significantly decrease and accuracy of OT baseline quickly collapse.
    }
    \label{fig:blending}
\end{figure}

    \paragraph*{Footskate Generated with Motion Blending}
        To push further our evaluation toward concrete applications, we additionally evaluate contact detection on motion sequences that have been \emph{blended}. Motion blending (or motion interpolation) is a well-known technique widely used in animation that consists in mixing existing motions with dynamic blending weights to create new motions. In our case, motion blending is interesting because it is known to easily produce footskate in resulting motions, which can then help us to evaluate contact detection on such perturbed motions. To do so, we randomly picked in our testset pairs of motion subsequences having identical foot contact patterns on either left or right foot. This approach enables us to make the hypothesis that foot contact patterns are preserved in blended motions, and thus serve as ground truth. We constituted a set of approximately $40'000$ blended motions of $80$ frames long (\ie $0.8s$) and quantified the amount of footskate introduced using the mean horizontal velocity of the feet during contact phases \cite{ref:SZKS19, ref:SZKZ20, ref:ZSKS18, ref:LZCv20}. We then evaluate once again foot contact detection using the $F_1$ score with these blended motions as inputs. We obtain the curves plotted in Figure~\ref{fig:blending} using simple moving average.
    
    As depicted in Figures~\ref{fig:noised}~to~\ref{fig:blending}, the proposed approach for foot contact labels detection is much more robust than threshold-based heuristics approaches represented by the OT baseline, regardless of the type of perturbation applied to motion sequences. Moreover, the improvement of modelling vGRFs instead of foot contact labels (Ours vs Ours-C) is much larger when facing perturbed motion sequences. Since vGRFs are much more related to motion dynamics than binary contact label, vGRFs estimation requires a deeper understanding of motion than contact labels detection. Indeed, approximately $90\%$ of contact labels can be correctly detected with foot position and velocity thresholding (see Table~\ref{tab:contacts}), \ie with almost no understanding. Then, in challenging conditions like perturbed input motion sequences, the performances of the variant modelling vGRFs are logically more stable since a deeper understanding of motion is intuitively more robust.

    \section{Footskate Cleanup}
        \label{section:FootskateCleanup}
In this section, we leverage our robust foot contact detection for a downstream task and propose a novel fully automatic workflow for footskate cleanup: the foot contact labels lead to kinematic constraints used in a dedicated optimisation-based inverse kinematics algorithm, where foot joints are constrained at ground contact points. The novelty of our approach is that the optimisation also includes the preservation of forces estimated by our network.

Let $\tilde{X}=(\tilde{Q}, S, \tilde{P})$ be the input motion sequence suffering from footskate, where $\tilde{Q}$ denote joint angles, $S$ the skeleton and $\tilde{P}$ the global trajectory positions of the root joint. Our goal is to find $Q$ and $P$ such that $X=(Q, S, P)$ is the footskate-cleaned version. First, we compute $\tilde{F}=\Psi_{\Theta}(FK(\tilde{Q}, S, \tilde{P}))$ and $C=\Gamma(\tilde{F})$, being respectively the vGRF components estimated by our deep neural network $\Psi_{\Theta}$ and the foot contact labels calculated using the contact function $\Gamma$ defined in Section \ref{section:Post-CaptureProcessing}. The function $FK(\cdot)$ refers to forward kinematics, i.e. the computation of joint positions from joint angles, skeleton and global trajectory. After initialising $Q$ and $P$ with $\tilde{Q}$ and $\tilde{P}$, we iteratively optimise both for a small fixed number $N$ of iterations to best satisfy foot contact constraints into a gradient-based optimisation loop. In the following paragraphs we describe each term $\mathcal{L}_i$ of the our loss function 
\begin{equation}\label{eq:footskateloss}
    \mathcal{L} = \omega_{q}\mathcal{L}_{quat} + \omega_{f}\mathcal{L}_{foot} + \omega_{t}\mathcal{L}_{traj} + \omega_{v}\mathcal{L}_{vGRFs}
\end{equation}
where weights $\omega_{i}$ are hyperparameters to balance our objectives.

\textbf{Quaternions loss $\mathcal{L}_{quat}$}: we parameterise joint angles with unit quaternions, which are well-suited for such an optimisation since they are free of singularities, computationally efficient, and numerically stable \cite{ref:MHLS21}. To keep valid rotations with quaternions throughout the optimisation, $\mathcal{L}_{quat}$ penalise quaternion norm deviations from~$1$:
\begin{equation}
    \mathcal{L}_{quat} = \sum_{t}{\sum_{j}{ \left( \left\lVert Q_j(t) \right\rVert - 1 \right)^2 }}
\end{equation}
where $Q_j(t)$ is the quaternion representing the orientation of joint $j$ at time $t$. 

\begin{figure}
    \centering
    \includegraphics[width=0.9\linewidth]{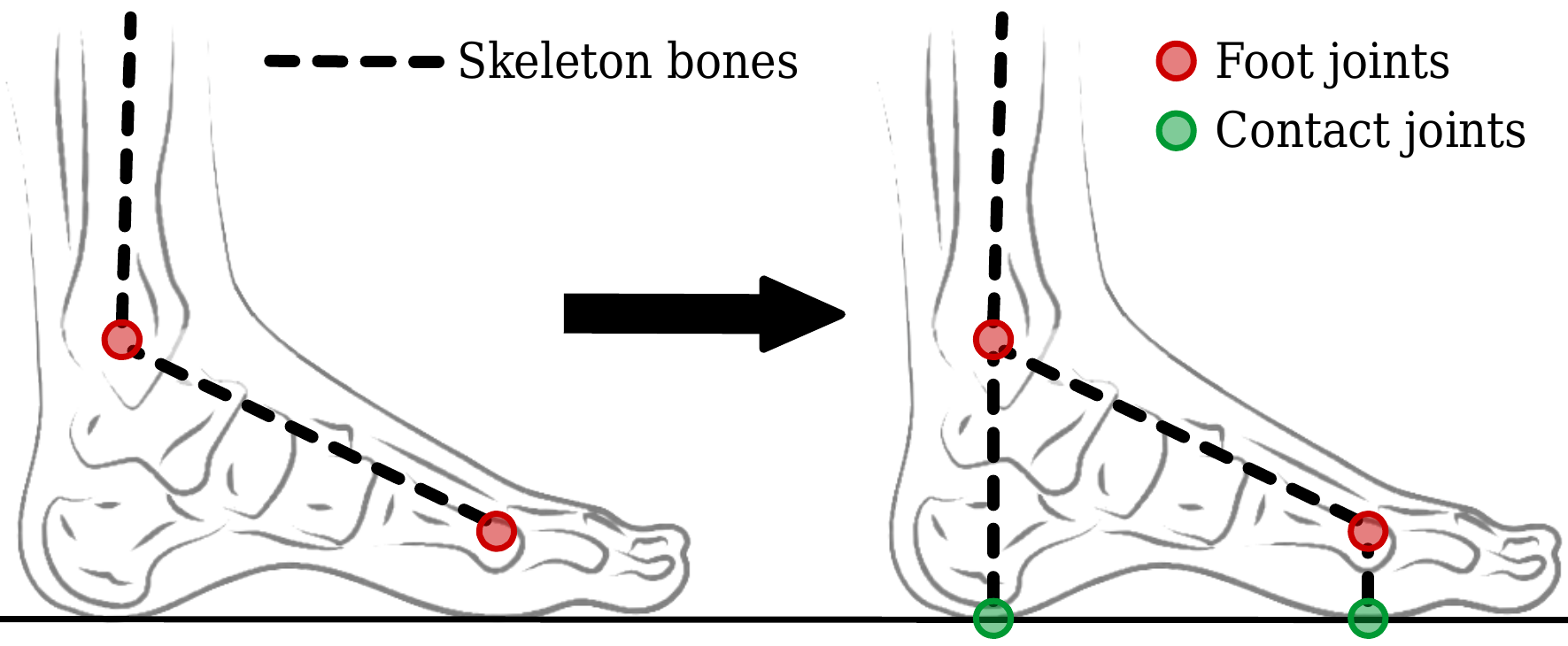}
    \caption{
        Illustration of \emph{contact joints} (green dots) inserted under corresponding \emph{foot joints} at the ground level in the skeleton to properly enforce foot contact constraints (See details below).
    }
    \label{fig:feet}
\end{figure}
\textbf{Foot contacts loss $\mathcal{L}_{foot}$}: satisfying foot contact constraints is not an easy task since we do not have directly access to actual contact locations w.r.t. the skeleton. Moreover, since \emph{foot joints} are located above actual contact points, they are allowed to rotate around the latter during part of contact phases (e.g. when the heel is in contact with the ground, the ankle is constrained to rotate around). To alleviate these issues, we artificially insert \emph{contact joints} in $S$ corresponding to contact points, \ie under foot joints at the ground level (see figure \ref{fig:feet}). We assume that each contact joint is located vertically under its foot joint in the T-pose at the ground height, has constant orientation w.r.t. its foot joint, and is assumed to be static during contact phases. Then, imposing zero velocity on contact joints instead of foot joints during the contact phases better reflects the kinematics of the interaction with the ground, allowing slight rotation of foot joints located above the ground level. To this end, we minimise the mean squared distance from contact joints to contact positions during contact phases to faithfully constrain foot joint positions through forward kinematics:
\begin{equation} \label{eq:loss_foot} 
    \mathcal{L}_{foot} = \sum_{c=1}^{C}{ \sum_{t=t_0(c)}^{t_1(c)}{ \Pi(t) \left\lVert FK(Q, S, P)_{j_c}(t) - p_c \right\rVert ^2 } }
\end{equation}
where $j_c$ and $p_c$ are respectively the joint in contact with the ground during contact phase $c$ and its contact position during the contact phase spanning from $t_0(c)$ to $t_1(c)$. Moreover, function $\Pi(\cdot)$ is a rectangular wave function with smoothed edges (synchronised with contact phases) to avoid sharp foot position changes.

\textbf{Trajectory loss $\mathcal{L}_{traj}$}: since foot contact constraints are not guaranteed to be reachable by leg extensions, the global trajectory might be affected. However, it is closely related to body dynamics and its optimisation must be carefully controlled to avoid artefacts. In particular, constraining trajectory positions might introduce implausible velocity changes, hurting motion dynamics realism, e.g. slowing down to reach some contact position might lead to speeding up afterwards to catch up the target trajectory. To this end, we minimise the trajectory velocity deviations instead of position deviations:
\begin{equation} \label{eq:loss_traj} 
    \mathcal{L}_{traj} = \sum_{t}{ \left( \lVert V(t) \rVert - \lVert \tilde{V}(t) \rVert \right)^2 },
    \quad
    V(t) = \frac{P(t + \Delta T) - P(t)}{\Delta T}
\end{equation}
where $V(t)$ is the global velocity of root joint at time $t$, computed from global trajectory position $P$ by numerical differentiation.

\textbf{vGRFs invariance loss $\mathcal{L}_{vGRFs}$}: the main novelty in our gradient-based optimisation approach is the use of our deep neural network through which gradients can flow. Since it is able to estimate consistent vGRF components from either clean or perturbed inputs, we further guide the optimisation by minimising the deviation between initially estimated vGRFs components $\tilde{F}$ and dynamically estimated vGRFs components $F=\Psi_{\Theta}(FK(Q, S, P))$:
\begin{equation}   
    \mathcal{L}_{vGRFs} = MSLE(F, \tilde{F})
\end{equation}
where MSLE is the mean squared logarithmic error, an alternative to mean squared error suited for vGRFs (see Section \ref{section:Method:TrainingAndInference}).

We used the same motion sequences blended for evaluation purposes in Section~\ref{section:Results:Perturbations} to test the proposed footskate cleanup workflow. With parameters $N=100$, $\omega_{q}=10^{-3}$, $\omega_{f}=10^{-5}$, $\omega_{t}=10^{2}$ and $\omega_{v}=5 \cdot 10^{-5}$ (see Equation~\ref{eq:footskateloss}), we achieve to significantly reduce the amount of footskate (the velocity of the feet during contact phases is approximately halved) while noticeably improving the realism of the sequences, assessed by visual inspection. We provide corresponding visual results of our footskate cleanup workflow in the supplementary video.
    \section{Conclusion}
        In this article, we proposed a novel approach that improves the state-of-art of foot contact detection and footskate cleanup in human character animation. Building on \UnderPressure, a novel motion capture database synchronised with pressure insoles which we release with this article, we proposed to learn the relationship between human motion and interactions between feet and ground with a deep neural network estimating vertical ground reaction forces, from which foot contact constraints can be derived and exploited for an effective removal of foot sliding artefacts.

As of today, the proposed database is one of a kind in animation as it provides synchronised motion capture and pressure insoles data for a wide variety of human motion. Our approach significantly outperforms thresholds-based heuristic approaches in detecting foot contact labels, which are limited in their capability to generalise. We show its robustness to perturbed input motion sequences, which is crucial in concrete use cases. We have shown that estimating forces greatly helps the accurate and robust detection of contacts. Finally, we demonstrate the usefulness of our foot contact detection approach for footskate cleanup, leveraging vGRF estimations to improve the quality of the results.

As described in Section \ref{section:RelatedWork}, foot contact labels are relevant to a lot of tasks in human animation and are typically obtained with simple heuristics, or manually annotated on an exceptional basis. Thus, many of these downstream tasks could probably benefit from the improved accuracy of our foot contact detection approach. Moreover, motion reconstruction relies more and more on modelling physical interaction of the feet with the ground (see Section \ref{section:RelatedWork:GRFs}). Resolving inverse dynamics problems like ground reaction forces estimation from motion capture data could provide a valuable regularization for such underconstrained problems as well as other tasks involving physics-based models. We believe our approach is a step in that direction.

    \printbibliography

    \clearpage

    \section{Supplementary Material}
        \subsection{CoP Prediction}
    In this section we provide results of our evaluation of Center of Pressure (CoP). Complementary to the root mean squared error (RMSE) of per-foot total vGRF which is mainly sensitive to global biases (at foot scale), the median absolute deviation (MAD) of the Center of Pressure (CoP) allows to evaluate predicted vGRF components with more sensitivy to local errors (at pressure cell scale) since CoP is calculated as the weighted mean of vGRF component application points. 
    
\begin{figure}
    \centering
    \includegraphics[width=0.9\linewidth]{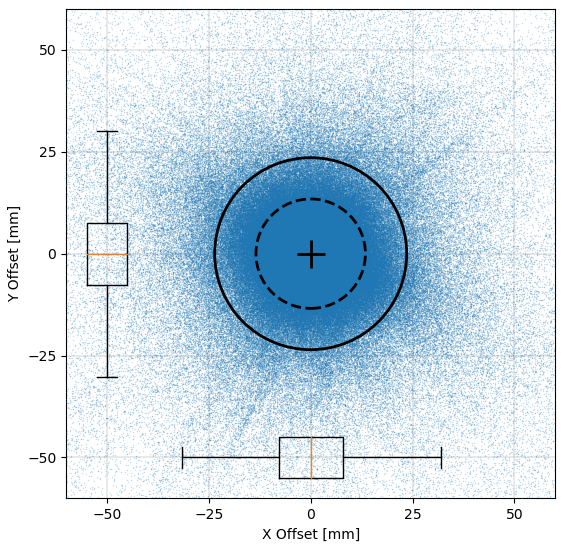}
    \caption{
        Scatter plot of 2D offsets between CoP computed from ground truth vGRFs and vGRFs predicted by our model. The concentric solid and dashed circles respectively represent the mean and median norm of 2D offsets, \ie one half of blue dots lie inside the dashed circle.
    }
    \label{fig:CoP_scatter}
\end{figure}

    Table~\ref{tab:CoP} provides the median absolute deviation (or median distance to ground truth) of the CoP, while Figure~\ref{fig:CoP_scatter} depicts the distribution of offsets from predicted to ground truth CoP.
    
    Similarly to our evaluation of vGRFs with RMSE, the results shown Table~\ref{tab:CoP} suggest that learning to model foot contact labels in addition to vGRF (Ours-C\&F) reduces accuracy, although the performances of both variants are close.

\vfill\null
\pagebreak

\renewcommand{\intercol}{\hskip 0.42cm}
\begin{table}
    \small
    \centering
    \caption{
        Median absolute deviation (MAD) in milimeters of CoP computed from estimated vGRF components. Similarly to estimated vGRF, modelling foot contact labels in addition to vGRFs (Ours) sligthly affects CoP accuracy.
    }

    \begin{tabular}{c | c @{\intercol} c @{\intercol} c @{\intercol} c @{\intercol} c @{\intercol} c | c}

        Model & \rotatebox{90}{Walking} & \rotatebox{90}{Running} & \rotatebox{90}{Obstacles} & \rotatebox{90}{Hopping} & \rotatebox{90}{Stairs} & \rotatebox{90}{Idle} & \rotatebox{90}{Overall} \\
        \hline

	    Ours-C\&F & 16.7 & 12.3 & 17.8 & 15.5 & 17.9 & 28.9 & 16.4 \\
	    Ours & \textbf{13.3} & \textbf{11.2} & \textbf{15.1} & \textbf{12.3} & \textbf{13.9} & \textbf{25.9} & \textbf{13.4}

    \end{tabular}
    \label{tab:CoP}
\end{table}

\end{document}